# Phonon-pair-driven Ferroelectricity Causes Costless Domain-walls and Bulk-boundary Duality

Hyun-Jae Lee†,[1], Kyoung-June Go†,[2], Pawan Kumar†,[1], Chang Hoon Kim[1], Yungyeom Kim[1], Kyoungjun Lee[3], Takao Shimizu[4], Seung Chul Chae[3], Hosub Jin[5], Minseong Lee[6], Umesh Waghmare[7], Si-Young Choi*,[2,8], Jun Hee Lee*,[1,9]

[1]Department of Energy Engineering, School of Energy and Chemical Engineering, Ulsan National Institute of Science and Technology (UNIST), Ulsan, 44919, Republic of Korea

[2]Department of Materials Science and Engineering, Pohang University of Science and Technology (POSTECH), Pohang, 37673, Republic of Korea

[3]Department of Physics Education, Seoul National University, Seoul, 08826, Republic of Korea

[4]Research Center for Electronic and Optic materials, National Institute for Materials Science, Tsukuba, 305-0044, Japan

[5]Department of Physics, Ulsan National Institute of Science and Technology (UNIST), Ulsan, 44919, Republic of Korea

[6]National High Magnetic Field Laboratory, Los Alamos National Laboratory, Los Alamos, New Mexico 87545, USA

[7]Theoretical Sciences Unit, Jawaharlal Nehru Centre for Advanced Scientific Research, Jakkur, Bangalore, 560064, India

[8]Department of Semiconductor Engineering POSTECH, Pohang 37673, Republic of Korea

[9]Graduate School of Semiconductor Materials and Devices Engineering, UNIST, Ulsan, 44919, Republic of Korea

† These authors equally contributed to this work.

**Summary**

Ferroelectric domain walls, recognized as distinct from the bulk in terms of symmetry, structure, and electronic properties, host exotic phenomena including conductive walls[1-3], ferroelectric vortices[4,5], novel topologies[6-8], and negative capacitance[9]. Contrary to conventional understanding, our study reveals that the structure of domain walls in $HfO_2$ closely resembles its bulk. First, our first-principles simulations unveil that the robust ferroelectricity is supported by bosonic pairing of all the anionic phonons in bulk $HfO_2$. Strikingly, the paired phonons strongly bond with each other and successfully reach the domain wall's centre without losing their integrity and produce bulk-like domain walls. We then confirmed preservation of the bulk phonon displacements and consequently full revival of the bulk structure at domain walls via aberration-corrected STEM. The newly found duality between the bulk and the domain wall sheds light on previously enigmatic properties such as zero-energy domain walls[10], perfect Ising-type polar ordering[10,11], and exceptionally robust ferroelectricity at the sub-nm scales[10,12]. The phonon-pairing discovered here is robust against physical boundaries such as domain walls and enables zero momentum and zero-energy cost local ferroelectric switching. This phenomenon demonstrated in Si-compatible ferroelectrics provides a novel technological platform where data storage on domain walls is as feasible as that within the domains, thereby expanding the potential for high-density data storage and advanced ferroelectric applications.

**Introduction**

Ferroelectric materials are characterized by a spontaneous electric polarization which can be switched by an external electric field. In oxide ferroelectrics, this polarization results from a polar phonon or ionic displacements requiring an essential volume referred to as a domain[13-16]. The interface where two domains of opposite polarization meet is known as the domain wall (DW). This region facilitates the transformation from one domain to another, either through gradual change, as observed in 180º Ising-type walls[17], or spatial rotation, as in Bloch/Néel-type walls[18,19]. At the DW, polarization is either reduced or completely eliminated; this deviation or fluctuation from the bulk incur a cost of finite domain wall energy (DWE).

In ferroelectrics, the structural changes at the DW often lead to novel properties. For instance, the conductivity of DWs in $BiFeO_3$ [1-3] has been explored for potential applications in memory device read mechanisms[20-22]. The concept of switchable ferroelectric vortices has been proposed for achieving high-density memory bits[4,5]. Energetically unstable head-to-head (or tail-to-tail) DWs in $(Y/Er)MnO_3$ are topologically fixed by the intersection of domain states[6-8]. DW studies have also delved into negative capacitance[9] which could reduce power needed in transistor switching. These phenomena are largely due to DWs being topological defects with distinct symmetries and electronic structures from the surrounding bulk material[23-25].

Contrary to conventional understanding, our research reveals the simplest type of DWs in $HfO_2$, whose structure has the identical symmetry with adjacent bulk domains. Diverging from typical ferroelectric behaviour where weakened condensation of polar phonons creates a distinct DW structure, DWs in HfO$_2$ exhibit a unique behaviour. Our study uncovers bosonic pairing of phonons—exceptionally rare in nature—as a critical factor stabilizing the ferroelectricity of HfO$_2$. This pairing enables the partner phonons in the pairs to mutually support each other across a domain wall, leading to a complete retention of the original bulk structure at the centre of the domain wall. This mutual support explains unconventional phenomena such as exceptionally narrow but robust sub-nanometre domains and the perfect Ising model behaviour where polarization at DWs does not diminish but results in degenerate mixed states of ferroelectric and antiferroelectric phases[11]. This complete reconstruction of bulk structure at the zero-cost DWs is directly visualized by our scanning transmission electron microscopy (STEM) measurements.

**Result and Discussion**

To explore the pairing relationship between O- and Hf-modes (**Figure S1**), which govern the phase transition from the paraelectric cubic $Fm\bar{3}m$ to the ferroelectric orthorhombic $Pca2_1$ structures in HfO$_2$, we analysed the lowest order trilinear couplings among these modes (**Figure 1**). We find that the $Y'_3$ mode, where Hf atoms are displacing in a type of antiphase along *y*-direction (**Figure 1**, dotted box) interacts with the pair of polar $\Gamma^z_{15}$ mode and anti-polar $Y^z_5$ mode, whose condensations notably suppress *z*-displacement of oxygen atoms in the closely spaced layers (referred to as the spacer layer, depicted as the shaded region) and amplifies their displacements in layers with wider spacing (termed the ferroelectric layer, shown as the unshaded region) (**Figure 1,** top). As a result, oxygen atoms in the ferroelectric layer predominantly contribute to the material's ferroelectricity while those in the spacer layer remain inert leading to an inherently patterned structure characteristic in the ferroelectric HfO$_2$. This phenomenon has been validated through both theoretical and experimental means[10,26]. Additionally, the Y$_3'$ mode couples trilinearly with (i) the pair of unstable X$_2'$ and stable Z$_5^x$ modes (**Figure 1**, middle), and (ii) the pair of X$_5^y$ and Z$_5'^y$ modes (**Figure 1**, bottom) and condensation of these pairs leads to reduction of the *x*- and *y*-displacement of oxygen atoms in the spacer layers while enhancement in the ferroelectric layers. In these paired phonons guaranteed by trilinear couplings, the oxygen modes Y$_5^z$, Z$_5^x$, and Z$_5'^y$ inherit additional momentum (π) along the *y*-direction relative to their partner modes Γ$_{15}^z$, X$_2'$, and X$_5^y$, respectively. Intriguingly, the partner phonons forming any of the three pairs are connected by a band running along y-direction of wavevector (**Figure 2c**).: Γ$_{15}^z$ links to Y$_5^z$ and X$_5^y$ to Z$_5'^y$ through a phonon band. While additional couplings at the W point cause splitting in the band connecting X$_2'$ and Z$_5^x$ modes, it does not affect the origin of the zero-cost DWs which are discussed in **Figure S2**.

While the Γ$_{15}^z$ & Y$_5^z$ modes, Z$_5^x$ & X$_2'$ modes, and Z$_5'^y$ & X$_5^y$ modes in HfO$_2$ are paired with each other, a typical perovskite ferroelectric such as BaTiO$_3$ exhibits a different mechanism in which its ferroelectric polarization is driven by the only Γ$_{15}^z$ mode. Here, the DWE is dictated by the single positive curvature of the polar phonon band, which accounts for the excitation of a phonon at the wavevector Γ+*q* (**Figure 2a**) and reflects in the disruption of translational symmetry at the domain wall separating domains of opposite polarization (−Γ$_{15}^z$) (**Figure 2b**). Consequently, formation of a domain wall in BaTiO$_3$ necessitates the absorption of phonons with nonzero momentum or structural fluctuation as depicted by the wavy lines (Γ→Γ+*q*) in the Feynman diagram (**Figure 2a**, below the phonon band) entailing an associated energy cost.

The unique phenomena of zero-energy and –momentum cost during DW formation in HfO$_2$ result from the momentum and energy compensation of the partner phonons in the pairs. For example, partner modes in the pair which are Y$_5^z$ and $\Gamma_{15}^z$ contributing equally to ferroelectricity belong to the same band but exhibit band curvatures of the opposite signs. The trilinear coupling between $\Gamma_{15}^z$ and Y$_5^z$ through Y$_3'$ phonon leads to a net zero momentum change along the *y*-direction during DW formation: the +*q* momentum from $\Gamma_{15}^z$ is precisely offset by the -*q* momentum from Y$_5^z$ (**Figure 2c**, as detailed in **Supplementary Note 1**). At the ferroelectric DW in HfO$_2$, the structural distortions corresponding to the two partner phonons reverse sign simultaneously, thereby nullifying their contributions to the energy cost. For instance, the energy cost of the deviation from $\Gamma_{15}^z$ phonon due to its momentum change (+*q*) is counterbalanced by the negative energy cost of the deviation from its partner phonon (Y$_5^z$) with the opposite momentum change (-*q*), as depicted by the dotted negative curvature in **Figure 2c**.

Similarly, the energy costs associated with the X$_2'$ and X$_5^y$ phonons at the wavevector (X+*q*) are compensated by those of Z$_5^x$ and Z$_5'^y$ modes at the wavevector (Z-*q*), respectively. This concurrent processes of energy excitation and compensation, with the perfect conservation of momentum, is illustrated in the Feynman diagram (**Figure 2c**). It depicts elastic phonon–phonon scattering with momentum transfer (*q*), a mechanism facilitated by the pairing of phonons. This interaction leads to a negligible energy cost for DW formation in HfO$_2$ (**Figure S3**).

Phonon pairing in HfO$_2$ is crucial in the restoration of the bulk structure at the domain wall. The DWs in HfO$_2$ are characterized by the simultaneous reversal of amplitudes of the partner modes involved in the pairing. For example, atomic displacements along the *z*-direction of the $\Gamma_{15}^z$ and $Y_5^z$ modes concurrently invert their sign to (-$\Gamma_{15}^z$ and -$Y_5^z$) across a DW (**Figure 2d**, top). Remarkably, the unit-cell-width structure from the centre of the DW, which separates domains with -$\Gamma_{15}^z$ and $\Gamma_{15}^z$ phonons, corresponds the structure of the $Y_5^z$ phonon (-$\Gamma_{15}^z|\Gamma_{15}^z \rightarrow Y_5^z$) locally, and vice versa (-$Y_5^z|Y_5^z \rightarrow$ -$\Gamma_{15}^z$). This mutual recovery of structural symmetry within the unit cell at the DW is effectively facilitated by the exchange of the partner phonons ($\Gamma_{15}^z \leftrightarrow Y_5^z$) in the pair. In a similar vein, the pairing partner phonons $X_2'$ and $Z_5^x$ recover the bulk structure at the domain wall by exchanging each other (-$X_2'|X_2' \rightarrow Z_5^x$ and -$Z_5^x|Z_5^x \rightarrow$ -$X_2'$ in **Figure 2d**, middle). Although $X_5^y$ and $Z_5'^y$ do not reverse their sign at the DW, remarkably they still display the same exchange relationship: -$X_5^y|X_5^y \rightarrow Z_5'^y$ and the -$Z_5'^y|Z_5'^y \rightarrow$ -$X_5^y$. This phenomenon where the bulk phonons do not diminish but exchange with their partner phonons underpins the bulk–boundary (domain wall) duality observed in $HfO_2$ and the details of the pairing direction of zone-boundary phonons is further illustrated in **Figure S4**.

To substantiate the conservation of bulk phonons at the DW in $HfO_2$, structural relaxations were conducted using density functional theory (DFT). These studies confirmed the consistent role of trilinear couplings (**Figure S5** and **Table S1**) among bulk phonon modes condensed in the ferroelectric phase. From this analysis, eight equivalent ferroelectric structures emerged. These include four up-polarized structures (U, $U_Y$, $U_X$, and $U_{XY}$) and four down-polarized structures ($D_{XY}$, $D_X$, $D_Y$, and D) (refer to **Figure S5–S6** for polarization switching results). For clarity, all phonon modes of the up-polarized structure U were assigned a positive sign. In contrast, some of the eight modes in other structures reverse their sign to maintain the sign consistency of their trilinear couplings as detailed in **Table S1**. Therefore, the eight equivalent structures, containing distinguishable sign combinations of the phonon modes, are labelled as (i) 'U' and 'D' according to the positive and negative sign of $\Gamma_{15}^z$, respectively, (ii) subscript 'X' according to the negative sign of $X_2'$, and (iii) subscript 'Y' according to the negative sign of $Y_5^z$. For example, the structural difference between U and $D_{XY}$ means not only $\Gamma_{15}^z$ mode changes the sign (U→D), but also $X_2'$ and $Y_5^z$ change the sign (subscript X and Y, respectively). Four distinct 180° ferroelectric DWs were identified: $D_{XY}$/U, $D_X$/U, $D_Y$/U, and D/U (**Figure S7–S8**). Among these, the $D_{XY}$/U DW was found to be the most stable astonishingly exhibiting almost zero DWE cost. In this DW structure, two pairs of the four phonon modes (the Γ-pair: $\Gamma_{15}^z$, $Y_5^z$ and the primary-mode pair: $X_2'$, $Z_5^x$) simultaneously reverse their sign across the DW (**Figure 3b**). During this switching, the absolute amplitudes of these modes are remarkably similar in bulk regions and the unit-cell of the DW (**Figure 3a**). As a result, the interfacial structure around the $D_{XY}$/U DW corresponds the bulk structure. Furthermore, the four modes $X_5^y$, $Z_5'^y$, $Z_5^x$, and $Y_3'$, including another pair ($X_5^y$ and $Z_5'^y$) maintain their sign and amplitudes nearly identical to the bulk structure during DW formation (**Figure 3a**). The absence of momentum change in the $D_{XY}$/U DW formation thereby precludes structural and energetic deviations from these four modes. This is further corroborated by the finding that the amplitudes and signs of all phonon modes at the unit-cell level at the DW are identical to those in the $D_X$ structure, one of the eight equivalent bulk structures (**Figure 3b**). This observation confirms that the interfacial region at the $D_{XY}$/U DW is almost indistinguishable from the bulk. In contrast, the pair-phonons in the other three domain walls ($D_X$/U, $D_Y$/U, and D/U) do not reverse their signs simultaneously, classifying these walls as energetically-costly conventional types (**Figure S7-S8**).

Our DFT findings were proved by experimental data obtained from annular bright field scanning transmission electron microscopy (ABF-STEM) measurements to verify the conservation of the phonon modes. The experimental data on *y*-directional distances between two oxygen atoms in spacer layers (represented as dumbbell-like structures, d$_y$, and indicated by black closed squares in the graph of **Figure 3c**) align closely with our DFT calculations (represented as open red squares in the same graph). The d$_y$ value is a summation of the displacements of the $X_5^y$ and $Z_5'^y$ phonon modes, which neither change their signs nor their structural consistency during D$_{XY}$/U DW formation. Therefore, the congruence between theoretical and experimental results for d$_y$ reinforces our predictions about the pairing and subsequent conservation of the coupled modes along the *y*-direction. Similarly, the *z*-directional off-centring displacements (d$_z$) in both spacer and ferroelectric layers, represented as black closed circles and red open circles respectively in Figure 3c, also exhibit good agreement with our DFT results, validating our hypothesis regarding the simultaneous sign reversal and conservation of the paired $\Gamma_{15}^z$ and $Y_5^z$ modes at the DW. Thus, the oxygen displacements around the DW, which closely resemble those in the surrounding bulk regions, confirm that the areas at and near the DW retain bulk-like structures. The experimentally measured d$_y$ and d$_z$ values and their comparisons between the other three DFT calculated DWs (D$_X$/U, D$_Y$/U, and D/U) are comprehensively summarized in **Figure S11**.

These experimental observations revealed an alternating arrangement of ferroelectric and spacer layers, accompanied by dumbbell-like oxygen positions (d$_y$ representing the displacement sum of the pair phonons $X_5^y$ and $Z_5'^y$) in the vicinity of DWs. This pattern aligns with our theoretical predictions indicating that the actual DWs comprise D$_{XY}$ and U structures. As illustrated in **Figure 3d–3f**, we note the emergence of exceptionally sharp domains characterized by alternating up/down polarization. Notably, a down-polarized domain, flanked by up-polarized domains, displayed an exceedingly narrow width of half a unit cell (**Figure**

**3d**). Additionally, the configuration featuring alternating down and up polarization regions was marked by distinct DWs (**Figure 3e**). Furthermore, **Figure 3f** depicts a more randomized pattern, corroborating our hypothesis that bulk-like structures with zero-cost proximal to $D_{XY}/U$ DWs are predominant in actual thin films in contrast to other high-energy DWs. The calculations of these bulk-like structures, along with the near-zero DWE of the observed configurations, are presented in the insets of **Figure 3d–3f**.

Our theoretical framework on phonon pairing not only predicts costless bulk-like DWs but also identifies DWs with higher energies (**Figure 4b, d**). We observed these costless DWs commonly appearing in ABF-STEM images of the thin film (**Figure 4a**), attributable to the zero-cost momentum and energy facilitated by the phonon pairs. Interpreting the top and right parts of **Figure 4a** as $D_{XY}$ and U, respectively, enables us to identify the DW where spacer layers come into contact as $U_Y/U$ (**Figure 4b**) and $D_X/D_{XY}$ (**Figure 4d**), both of which exhibit the second lowest DWE according to our DFT results (**Figure S9**). Although the white dashed region is expected to be a junction between nonpolar spacer layers (S S), the measured oxygen displacements (0.112Å, 0.081Å in **Figure 4b** and **4d**) are smaller than those in the bulk ferroelectric layers but comparable to our DFT calculations (0.125Å in **Figure 4c** and **4e**).

We attribute the meta-stability of these second-lowest energy DWs to the concept of "Anti-pairing" where only one of the partner phonons in each pair changes sign at the DW (**Figure 4a**). This phenomenon, wherein $Y_5^z$ and $Z_5^x$ modes change their sign across the DW but their partner phonons ($\Gamma_{15}^z$ and $X_2'$, respectively) do not, incurs energy cost and thus does not produce the bulk-like DW. The sign change in amplitudes of phonons in the anti-pairing DWs provides insight into the rest of the DWs. Defining the two anti-pairing DWs as $U_Y/U$ and $D_X/D_{XY}$, the bottom left of **Figure 4a** can be interpreted as $U_Y/D_X$, which is another zero-cost DW (across which the signs of $\Gamma_{15}^z$, $X_2'$, $Y_5^z$, and $Z_5^x$ modes change). Our phonon-based analysis thus serves as a powerful tool for elucidating the symmetry and stability of various types of DWs.

While fermionic particles such as electrons are known to pair and induce observable effects such as superconductivity, bosonic entities such as phonons rarely exhibit pairing. Our research demonstrates how the condensation of bound pairs of phonons, facilitated through interactions with a mediating phonon, leads to the emergence of unusual ferroelectricity and a duality (equivalence) between the ferroelectric bulk and its domain walls in $HfO_2$. This duality manifests in the formation of atomic-scale domains separated by ultra-thin walls, resembling the bulk crystal, and characterized by zero energy and momentum cost. These findings advance our understanding of how phonon interactions result in bosonic pairing, illuminating a relatively unexplored phenomenon. Furthermore, the insights gleaned from this study open up promising avenues for future research, particularly in the design and development of ultimately-dense ferroelectric memories compatible with silicon devices.

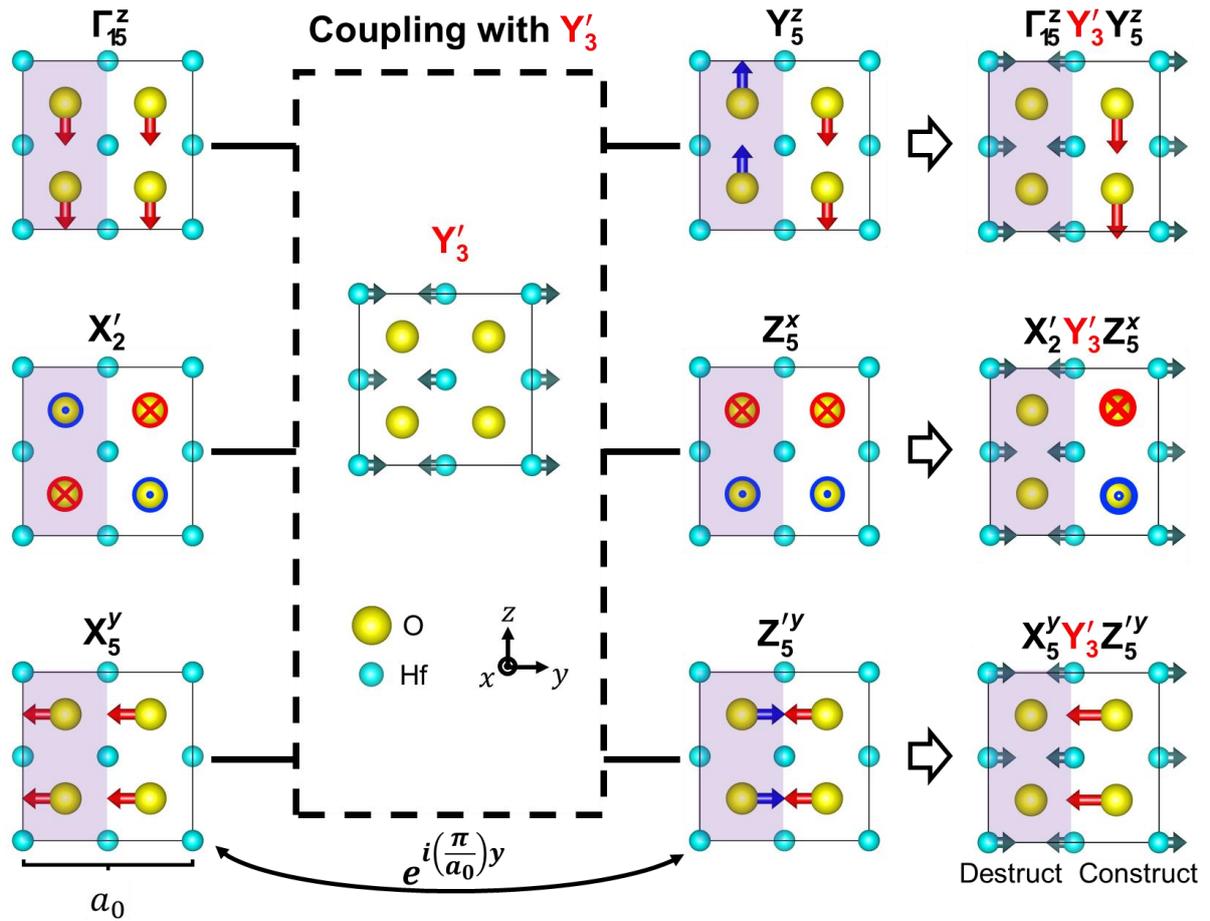

**Figure 1 | Phonon pairs by coupling with $Y_3'$ mode in fluorite structure**. The A-type antipolar $Y_3'$ mode along the *y*-axis gives *y*-directional momentum (**dotted box**) to other modes through trilinear couplings, resulting in the formation of phonon pairs in the ferroelectric (Pca2$_1$) HfO$_2$. This includes the pair of $\Gamma_{15}^z$ (polar mode) and $Y_5^z$ modes with *z*-directional oxygen displacements (**top**), as well as the pair of $X_2'$ (soft mode) and $Z_5^x$ modes with *x*-directional oxygen displacements (**middle**), and the pair of $X_5^y$ and $Z_5'^y$ modes with the *y*-directional oxygen displacements (**bottom**). These three pairs with Cartesian directional (*x*, *y* and *z*) oxygen displacements exist in the ferroelectric HfO$_2$ simultaneously resulting in offset oxygen displacements in narrow Hf spacing (shaded) region.

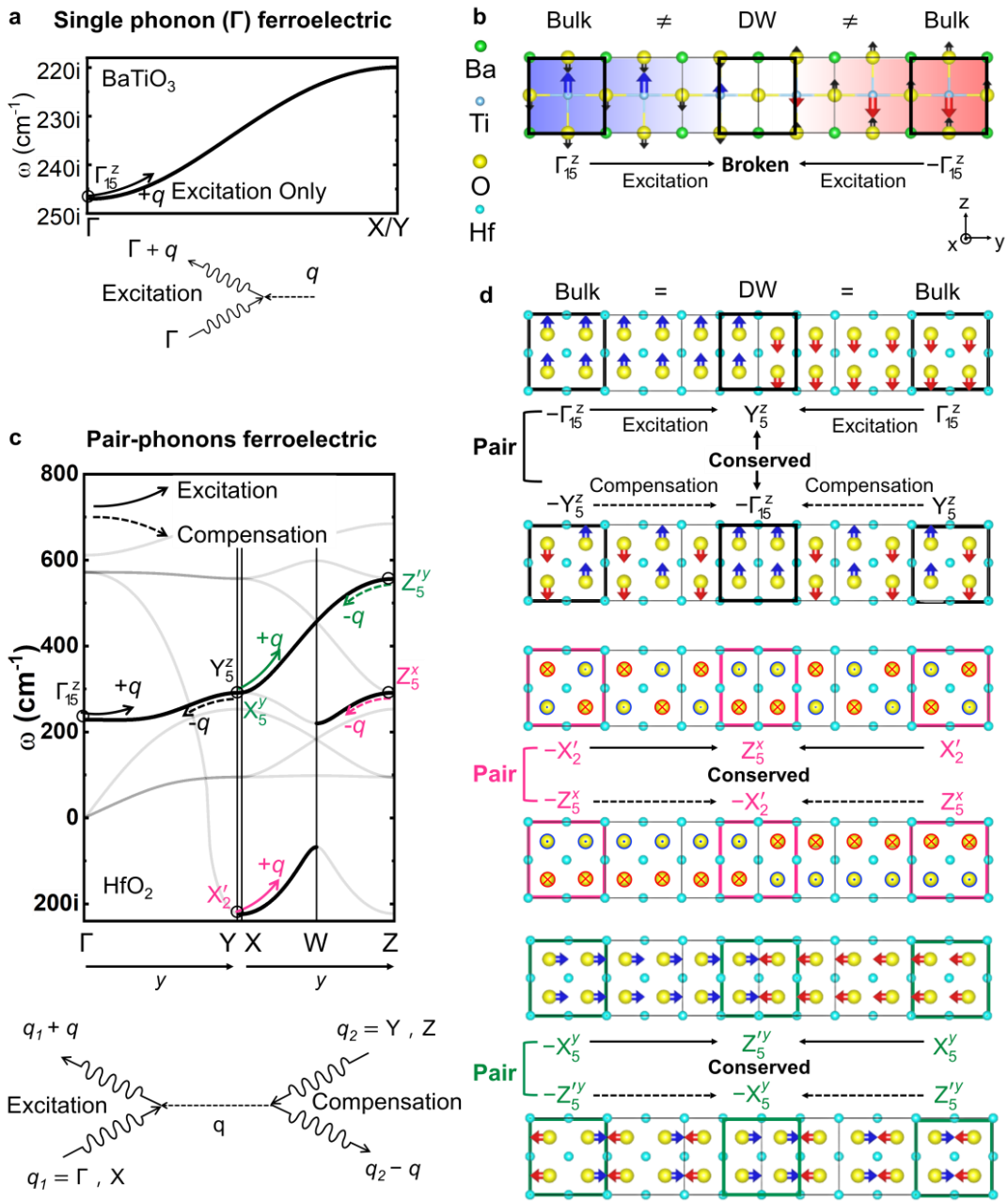

**Figure 2 | A comparison between single phonon ferroelectrics and Pair-phonon ferroelectrics in 180° domain walls.** In single-phonon ferroelectrics like BaTiO$_3$ where the paraelectric to ferroelectric phase transition is driven by a single unstable polar $\mathbf{\Gamma_{15}^z}$ mode, alone (a, top), the momentum change (+q) in 180º DW formation excites this mode (a, bottom) resulting in an energy cost that can be estimated by the curvature (indicated by a solid curved

arrow) of the band along the DW direction (a, top). Atomic displacements in the two adjacent unit-cells are canceled out leading to suppressed amplitude of the $\mathbf{\Gamma_{15}^z}$ mode near the domain wall, making its structure deviate from the bulk regions (b) in such ferroelectrics. In contrast, pair-phonon ferroelectrics like HfO$_2$ exhibit a paraelectric to ferroelectric phase transition involving pair-phonons (c, top). The energy excitation due to the momentum change (+q) of phonon modes ($\mathbf{\Gamma_{15}^z, X_2', X_5^y}$) is compensated by the momentum reduction (-q) of their respective band partner phonons ($\mathbf{Y_5^z, Z_5^x}$ and $\mathbf{Z_5'^y}$) (c, bottom) resulting in zero momentum change and zero energy cost in domain wall formation in such ferroelectrics. The atomic displacements associated with eigenvectors of these modes in the interfacial region around the domain wall are fully conserved and recovered by their pair-phonons (d) which make the interfacial unit cells indistinguishable from bulk.

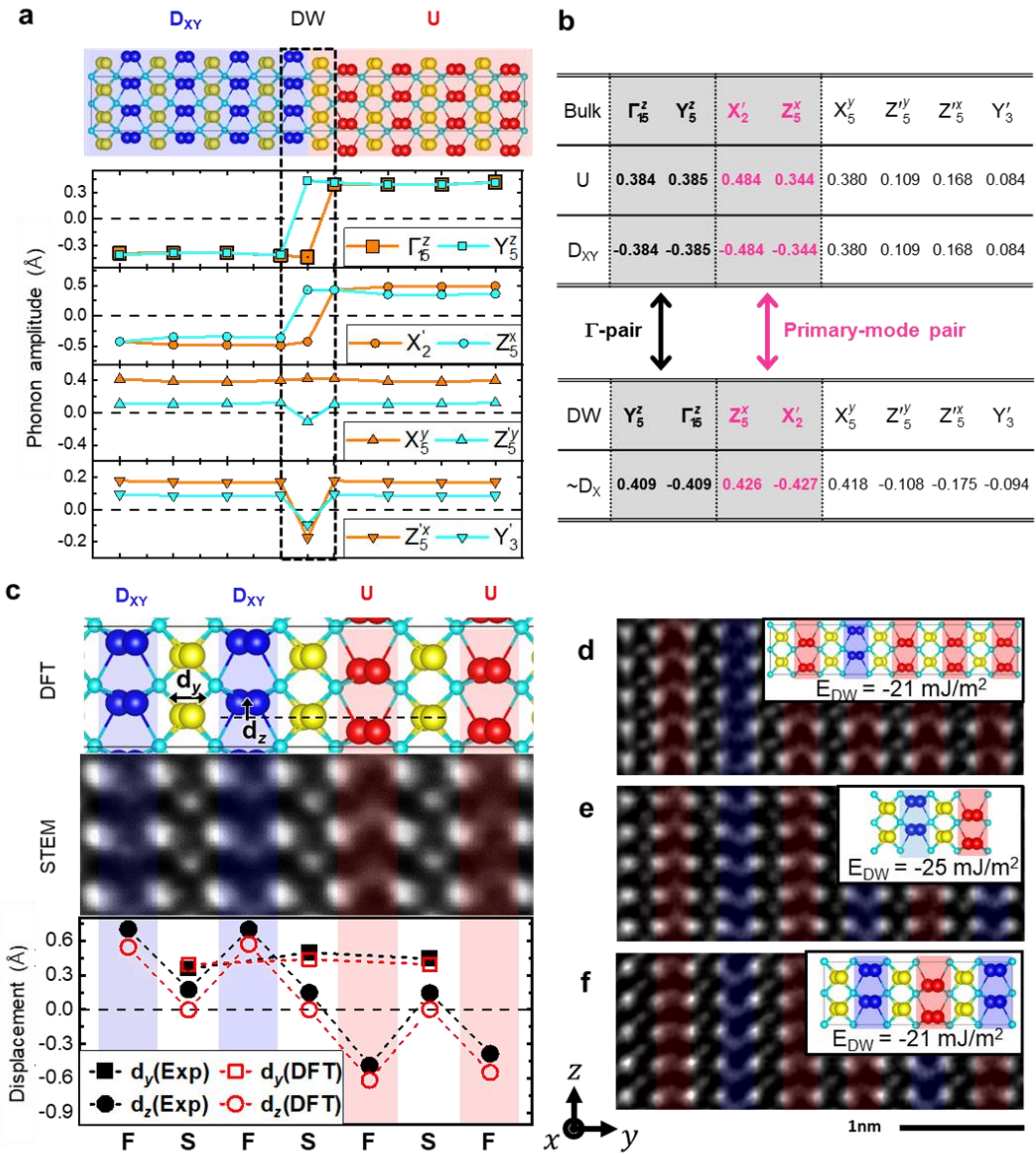

**Figure 3 | Conserved local phonon displacements and perfect Ising ordering verified with the consistency between experimental measurements and DFT calculations.** Atomic structure of the lowest DWE $D_{XY}$/U domain consists of four unit cells of down-polarized $D_{XY}$ (left blue) and four up-polarized U unit cells (right red) and the interfacial region of DW marked by black dashed lines (**a,** top). The mode amplitudes for each sub-unit cell (**a,** bottom) demonstrating that across the DW, the amplitudes of each local phonon mode remain constant despite changes in their signs. Interfacial regions around the DW contain local phonon modes

with the same sorts and similar amplitudes as the mono-domain structure $D_X$ (**b**). The theoretical and experimental atomic displacements of the structure consisting of two $D_{XY}$ and two U unit cells are compared (**c**), which show nearly identical *y*-direction ($d_y$) and *z*-direction ($d_z$) distances between two oxygen atoms in the spacer (yellow atoms) and ferroelectric (blue atoms) layers, respectively. F and S denote the ferroelectric and spacer layers respectively. Annular bright field scanning transmission electron microscopy (ABF-STEM) images of various DW configurations (**d**)-(**f**), including the structure consisting of a single sub-unit cell with opposite polarization from the surrounding domain (**d**), sharp alternating domains with opposite polarizations (**e**) and randomly mixed domains (**f**). The DWEs for each configuration were shown as insets of (**d**)-(**f**). The intensity of ABF-STEM images is inverted. Colored columns represent the polarization direction: red for up and blue for down.

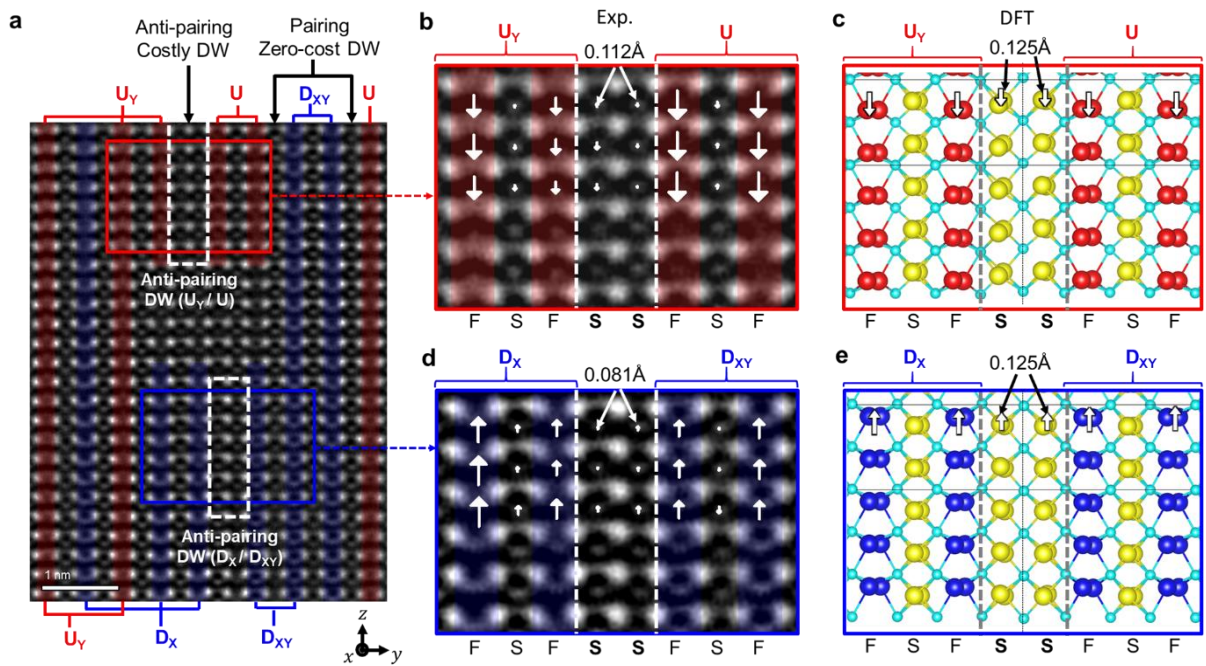

**Figure 4 | Costly domain walls by anti-paired phonons in contrast to zero-cost domain wall by paired phonon.** An inverted annular bright field scanning transmission electron microscopy (ABF-STEM) image reveals alternating pattern of spacer (S) and ferroelectric (F) layers, except for the regions outlined by red and blue boxes confirming the presence of zero-cost domain walls (**a**). In the zoomed red box region, oxygen displacement mapping clearly shows DWs formed between spacer layers with subtle displacement, indicated by white dashed lines (**b**). White arrows highlight the oxygen displacement away from the geometric centre of neighbouring hafnium atoms, opposite to the polarization direction. DFT calculations confirmed the structure to be a DW between $U_Y$ and U domains (**c**). In the zoomed blue box region, the oxygen displacement mapping reveals another type of unconventional domain wall formed by spacer layers (**d**). This was again confirmed by DFT calculations of the DW between $D_X$ and $D_{XY}$ structures (**e**). The anti-pairing in the DW between U and $U_Y$ ($D_X$ and $D_{XY}$), denoted by white dashed line boxes in (**a**, **b** and **d**), results in the reversal of only $Y_5^z$ and $Z_5^x$ phonons without any pairings across the DW.

**Acknowledgments:** We thank S. Rhim for the scientific discussions. This work was supported by the Next-generation Intelligence Semiconductor R&D Program (2022M3F3A2A01079710), Creative Materials Discovery (2017M3D1A1040828),

Midcareer Researcher (2020R1A2C2103126), Basic Research Laboratory (RS-2023-00218799), Nano & Material Technology Development Program (RS-2023-00260171), 2021R1I1A1A01057760 and RS-2023-00257666 through the National Research Foundation of Korea (NRF) funded by the Korea government (MSIT). This work was also supported by the Korea Institute for Advancement of Technology (KIAT) grant funded by the Korea Government (MOTIE) (P0023703, HRD Program for Industrial Innovation) and the National Supercomputing Center with supercomputing resources including technical support (KSC-2023-CRE-0547, KSC-2022-CRE-0076, KSC-2022-CRE-0454, KSC-2022-CRE-0456). SYC acknowledges the support by National R&D Program through the National Research Foundation of Korea (NRF) funded by Ministry of Science and ICT(RS-2023-00258227). UVW acknowledges support from a JC Bose National Fellowship of SERB-DST, Government of India. A portion of this work was performed at the National High Magnetic Field Laboratory, which is supported by National Science Foundation Cooperative Agreement No. DMR-2128556* and the State of Florida and the U.S. Department of Energy.

**Competing Interests:** There are no conflicts of interest to declare.

**Author contributions:** JHL supervised the work. HJL and JHL carried out DFT calculations, and CHK, ML and YK helped analysing the data. KJG and SYC provided experimental verifications by STEM. KL, TS and SCC conducted thin film growth. JHL, HJ, PK and UVW developed the theory. JHL, HJL, UVW and PK wrote the manuscript.

**Data and materials availability:** All data is available in the main text or the supplementary materials.